\documentclass[a4paper]{IEEEtran}
\IEEEoverridecommandlockouts
\usepackage{cite}
\usepackage{amsmath,amssymb,amsfonts}
\usepackage{algorithmic}
\usepackage{graphicx}
\usepackage{textcomp}
\usepackage{xcolor}
\usepackage{acronym}
\usepackage{import}
\usepackage{tikz}
\usepackage{siunitx}
\usepackage{url}
\usepackage{multirow}
\usepackage[super]{nth}
\usepackage{rotating}
\usepackage{subcaption}
\usepackage{makecell}
\usepackage{tikz}

\usepackage{hyperref}

\usetikzlibrary{positioning}
\newcommand\tabrotate[1]{\begin{turn}{90}\rlap{#1}\end{turn}}
\def\BibTeX{{\rm B\kern-.05em{\sc i\kern-.025em b}\kern-.08em
    T\kern-.1667em\lower.7ex\hbox{E}\kern-.125emX}}

\def\vspaceImage{-5mm}

\begin{document}

\newcommand\copyrighttext{%
	\footnotesize \textcopyright 2020 IEEE. Personal use of this material is permitted.
	Permission from IEEE must be obtained for all other uses, in any current or future 
	media, including reprinting/republishing this material for advertising or promotional 
	purposes, creating new collective works, for resale or redistribution to servers or 
	lists, or reuse of any copyrighted component of this work in other works. 
	DOI: \href{https://doi.org/10.1109/QoMEX48832.2020.9123140}{10.1109/QoMEX48832.2020.9123140} }
\newcommand\copyrightnoticeOwn{%
	\begin{tikzpicture}[remember picture,overlay]
		\node[anchor=north,yshift=-10pt] at (current page.north) {\fbox{\parbox{\dimexpr\textwidth-\fboxsep-\fboxrule\relax}{\copyrighttext}}};
	\end{tikzpicture}%
	\vspace{-8mm}
}

\title{On Versatile Video Coding at UHD with Machine-Learning-Based Super-Resolution\\}


\author{\IEEEauthorblockN{Kristian Fischer, Christian Herglotz, and Andr\'e Kaup\\}
	\IEEEauthorblockA{\textit{Multimedia Communications and Signal Processing} \\
		\textit{Friedrich-Alexander-Universit\"at Erlangen-N\"urnberg (FAU)}\\
		Cauerstr. 7, 91058 Erlangen, Germany \\
		\{Kristian.Fischer, Christian.Herglotz, Andre.Kaup\}@fau.de}
}
\maketitle 
\copyrightnoticeOwn

\begin{abstract}
	Coding 4K data has become of vital interest in recent years, since the amount of 4K data is significantly increasing. We propose a coding chain with spatial down- and upscaling that combines the next-generation VVC codec with machine learning based single image super-resolution algorithms for 4K. The investigated coding chain, which spatially downscales the 4K data before coding, shows superior quality than the conventional VVC reference software for low bitrate scenarios. Throughout several tests, we find that up to 12~\% and 18~\% Bj\o ntegaard delta rate gains can be achieved on average when coding 4K sequences with VVC and QP values above 34 and 42, respectively. Additionally, the investigated scenario with up- and downscaling helps to reduce the loss of details and compression artifacts, as it is shown in a visual example.
\end{abstract}



\section{Introduction}

The amount of multimedia data in Ultra-High-Definition~(UHD) resolution which is consumed all over the world is rapidly increasing. According to the Cisco Visual Networking Index \cite{cisco2019}, the percentage of installed flat panel TVs that are able to show Videos in UHD resolution will be increasing from 23 \% in 2017 to 62 \% in 2022. Derived from this statistic, it can be seen that the significance of UHD is growing such that the provision of suitable data is essential.

Hence, the question arises how to compress these massive amounts of data in the future, such that the end user can watch UHD content with the best quality at reasonable bitrates. For transmission scenarios with low available bitrates, it has recently been shown \cite{bruckstein2003,hamis2019, nguyen2008,shen2011,georgis2016} that it is beneficial to downscale the video first before compressing it into the bitstream.

\begin{figure}[t]
	\centering
	\resizebox{0.49\textwidth}{!}{
		\begin{tikzpicture}[]

\def\height{0.5}
\def\width{0.5}
\def\offsetBranches{-4}
\def\letterSize{\large}
\def\textWidth{\width *2 cm -0.2cm}
\def\offsetBoxes{1.0}

\tikzset{>=latex}
\node[](input) at (-0.0, -2){\letterSize{\makecell[c]{Original\\Frame\\$X_\mathrm{orig}$}}};


\draw (2*\offsetBoxes - \width,\height) rectangle (2*\offsetBoxes + \width,-\height) node[pos=.5, text width=\textWidth,align=center] (down){\letterSize$\downarrow s$};

\draw (4*\offsetBoxes - \width,\height) rectangle (4*\offsetBoxes + \width,-\height) node[pos=.5, text width=\textWidth,align=center] (enc){\letterSize Enc.};

\draw (6*\offsetBoxes - \width,\height) rectangle (6*\offsetBoxes + \width,-\height) node[pos=.5, text width=\textWidth,align=center] (dec){\letterSize Dec.};

\draw (8*\offsetBoxes - \width,\height) rectangle (8*\offsetBoxes + \width,-\height) node[pos=.5, text width=\textWidth,align=center] (up){\letterSize$\uparrow s$};	
	
\node[](sequence1) at (10*\offsetBoxes,0){\letterSize \makecell[c]{Decoded\\Frame\\$\tilde{X}_\mathrm{scaled}$}};

\draw (4*\offsetBoxes - \width, \offsetBranches+\height) rectangle (4*\offsetBoxes + \width,\offsetBranches-\height) node[pos=.5, text width=\textWidth,align=center] (enc2){\letterSize Enc.};

\draw (6*\offsetBoxes - \width, \offsetBranches+\height) rectangle (6*\offsetBoxes + \width,\offsetBranches-\height) node[pos=.5, text width=\textWidth,align=center] (dec2){\letterSize Dec.};

\node[](sequence2) at (10*\offsetBoxes,\offsetBranches){\letterSize \makecell[c]{Decoded\\Frame\\$\tilde{X}_\mathrm{conv}$}};

\draw[->, line width=0.5mm] ([xshift=-0.14cm]input.east)-- ++(0.25, 0) |- (down.west);
\draw[->, line width=0.5mm](down)--node[above]{\letterSize$X_\mathrm{LR}$}(enc) ;
\draw[->, line width=0.5mm](enc)--node[above]{\letterSize$b_\mathrm{LR}$}(dec);
\draw[->, line width=0.5mm](dec)--node[above]{\letterSize$\tilde{X}_\mathrm{LR}$}(up);
\draw[->, line width=0.5mm](up)--(10*\offsetBoxes - \width - 0.2,0);

\draw[->, line width=0.5mm] ([xshift=-0.14cm]input.east) -- ++(0.25, 0) |- (enc2.west);
\draw[->, line width=0.5mm](enc2)--node[above]{\letterSize$b_\mathrm{HR}$}(dec2);
\draw[->, line width=0.5mm](dec2)--(10*\offsetBoxes - \width - 0.2,\offsetBranches);
\end{tikzpicture}
	}
	\caption{Top branch: coding chain with spatial up- and downscaling; Bottom branch: conventional coding}
	\label{fig:coding chain}
	\vspace{\vspaceImage}
\end{figure}
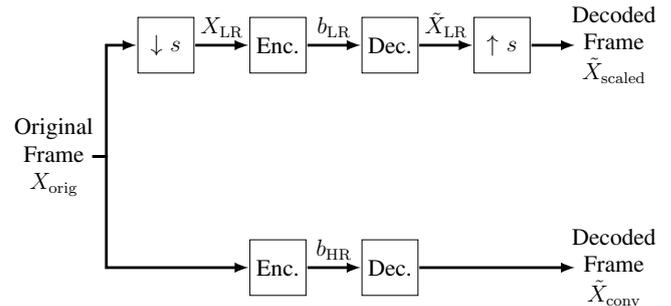

In Fig.~\ref{fig:coding chain}, such a coding chain with spatial up- and downscaling is depicted. First, the original frame $X_\mathrm{orig}$ is downscaled by a scaling factor $s$ before encoding. The resulting low resolution frame $X_\mathrm{LR}$ has $\frac{1}{s^2}$ pixels of the original $X_\mathrm{orig}$. At the decoder side, the bitstream $b_\mathrm{LR}$ is decoded into the frame with reduced resolution and lower quality $\tilde{X}_\mathrm{LR}$. Finally, the compressed low resolution frame is upscaled to the original resolution which results in $\tilde{X}_\mathrm{scaled}$. The lower coding branch in Fig.~\ref{fig:coding chain} shows the conventional coding process without modifying the spatial resolution.


Already in 2003, Bruckstein et al. \cite{bruckstein2003} proposed a coding chain in which they downscaled still images before encoding them with the JPEG standard. This method resulted in superior quality at low bitrates. 
In a similar approach \cite{hamis2019}, it was investigated using a generative adversarial network (GAN) to regain the spatial resolution of the downscaled images that were stored in BPG file format.

Nguyen et al. \cite{nguyen2008} showed for video data that it is beneficial to use a coding chain with an adaptive down-/up-sampling coding scheme at low bitrates for MPEG-2 encoding. There, the rate-distortion performance of the coding chain with down- and upsampling is superior to regular coding up to a certain bitrate which they called ``critical bitrate". 

For the H.264/AVC, Shen et al. \cite{shen2011} claimed that a coding chain with down- and upsampling diminishes blocking effects at low bitrates, since the number of discrete cosine transformation coefficients is reduced. Consequently, more bits can be spent for the remaining coefficients.

Using 4K video data, the authors in \cite{georgis2016} introduced a coding chain with spatial scaling in combination with the H.265/HEVC standard. There, they showed that the critical bitrate, up to which it is beneficial to use the coding chain with down- and upscaling, lies around 3.5 Mbit/s for their upscaling algorithm called Low-complexity Statistical Edge-Adaptive Back-projected Interpolation (L-SEABI). The main idea of this upscaling algorithm is to iteratively enhance the resolution of the single frames with a mixture of different bicubic filters.

\begin{figure}[t]
	\centering
	\resizebox{0.49\textwidth}{!}{
		\import{Tikz_vdsr/}{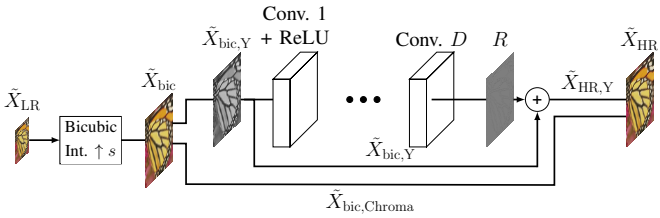}
	}
	\caption{Structure of the VDSR network enhancing the quality of a downscaled frame $\tilde{X}_\mathrm{LR}$ to a high resolution image $\tilde{X}_\mathrm{HR}$.}
	\label{fig:vdsr structure}
	\vspace{\vspaceImage}
\end{figure}


With the advent of the new Versatile Video Coding (VVC) codec \cite{segall2017} and motivated from the previously mentioned approaches, it is of vital interest to test the coding chain with spatial up- and downsampling also for this new codec. To regain the original resolution at the decoder side, two neural super-resolution networks are investigated. On the one hand, the Very-Deep Super-Resolution Network (VDSR) \cite{kim2016} is slightly modified and placed into the investigated coding chain. On the other hand, the more recent Residual Dense Network~(RDN)~\cite{zhang2018_RDN} is used to regain the original resolution. To evaluate this framework with different upscaling procedures, the Peak-Signal-to-Noise-Ratio (PSNR) and the Video Multi-Method Assessment Fusion (VMAF) metric~\cite{li2016_vmaf} are taken into account to generate rate-distortion curves for video sequences with 4K resolution ($3840\times2160$ pixels). As a reference, the coding chain is also tested with the L-SEABI upscaling algorithm proposed by Georgis et al. \cite{georgis2016}, as well as bicubic interpolation.


The remaining sections are organized as follows: In the subsequent Sec.~\ref{sec:basic algorithms}, the main components of the investigated framework are described. Subsequently, the simulation framework itself and the used neural super-resolution implementations are explained in Sec.~\ref{sec:coding chain}. The evaluation results are given in Sec.~\ref{sec:experimental results}. Finally, Sec.~\ref{sec:conclusion} concludes the paper and future plans are briefly discussed.

\section{Basic Algorithms}
\label{sec:basic algorithms}

Inspired by previous approaches, the investigated framework combines the next generation video codec VVC with the already existing single image super-resolution networks VDSR and RDN. Thus, these components are introduced in the following section.

\subsection{VVC Standard}

VVC is planned to be the next video compression standard by the Joint Video Experts Team (JVET). A first version of the new standard is expected at the end of 2020 \cite{segall2017}. VVC is supposed to save around 50 \% bitrate while maintaining the same visual quality compared to its predecessor HEVC \cite{sullivan2017}. VVC not only targets high resolution content but also modern video formats like $\ang{360}$, high dynamic range, and screen or medical content \cite{chen2019}.

One major difference from VVC to HEVC is that the maximum Coding Tree Unit (CTU) size is doubled to $128\times128$ pixels. Besides, VVC can also decide for ternary and non-square splits in the quadtree structure in order to better adapt to the video content.


Additional new features, apart from the block structure, are an extended number of intra prediction modes, an affine motion compensated prediction, improved entropy coding, and other minor extensions. Further details can be found in \cite{chen2019}.

\subsection{VDSR Network}
Since the upscaling at the decoder side is fundamental for the resulting quality of the investigated coding framework, VDSR is chosen because of the decent upscaling quality and the simple network structure. Besides, VDSR is only applied to the Y channel of an image or frame which makes it fit perfectly into a compression scheme with 4:2:0 YCbCr color sub-sampling. The two chroma channels are upscaled with bicubic interpolation.

VDSR is a very deep convolutional neural network for enhancing the resolution of single images proposed by Kim et al. in \cite{kim2016} and uses an end-to-end mapping between low and high resolution patches. The structure of the network, as it is placed into the coding chain, can be seen in Fig.~\ref{fig:vdsr structure}. First, the low resolution image $\tilde{X}_\mathrm{LR}$ is upscaled to the desired output resolution $W\times H$ with bicubic interpolation. The resulting bicubically interpolated image $\tilde{X}_\mathrm{bic}$ already contains the low frequency components of the image, but lacks the high frequencies which results in a blurry image. Afterwards, the Y\nobreakdash-channel of the bicubically interpolated image $\tilde{X}_\mathrm{bic,Y}$ is taken as an input of the deep convolutional neural network which estimates a residual image $R$ that contains the missing high frequency components of $\tilde{X}_\mathrm{bic,Y}$. The Y-channel of the final output image $\tilde{X}_\mathrm{HR,Y}$ is calculated by adding the input and the output of VDSR network with
\begin{equation}
\tilde{X}_\mathrm{HR,Y} = \tilde{X}_\mathrm{bic,Y} + R.
\end{equation}

All in all, VDSR network consists of $D=20$ convolutional layers which are all followed by a Rectified Linear Unit (ReLU) activation layer that maps all negative values to zero, except the last layer. In the first layer, the bicubically interpolated image $\tilde{X}_\mathrm{Y, bic}$ of the size $W\times H$ is convolved with 64 filters of the size $1\times3\times3$ which results in 64 feature map channels, where each channel has the resolution $W\times H$. Subsequently, every convolutional layer consists of 64 filters of size $64\times3\times3$. The \nth{20} and final layer only consists of one filter with the size $64\times3\times3$ which results in the output channel $R$. Summarizing, it can be said that the actual increase in resolution comes from the bicubic interpolation and VDSR network enhances this blurry image with multiple trained filters. 


%

%

For faster inference and to train VDSR with high resolution images, we implemented and trained VDSR in Tensorflow~\cite{tensorflow2015-whitepaper} and Keras~\cite{chollet2015keras}, which is described in Sec. \ref{subsec: upscaling with VDSR}.
\subsection{RDN}
\label{subsec:RDN}

In addition to VDSR, a more recent super-resolution network is selected with the RDN by Zhang et al.~\cite{zhang2018_RDN}. Contrary to VDSR, no bicubic upscaling step is needed. Instead, hierarchical features are extracted by so-called residual dense blocks (RDBs) and from the resulting features, an upscaling network is used to increase the spatial resolution. Furthermore, RDN takes an RGB image as input and upscales all three color channels. Instead of VDSR, it is not applicable in a coding chain without color conversions from YCbCr to RGB and vice versa.

\begin{figure}[t]
	\centering
	\includegraphics[width=1\linewidth]{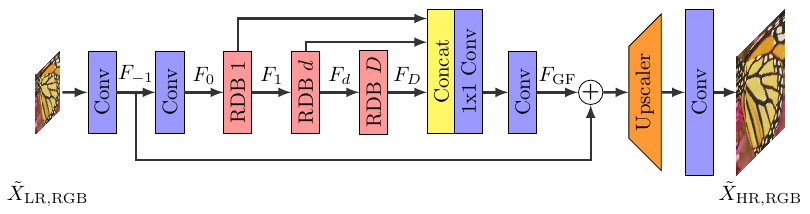}
	\caption{RDN architecture from \cite{zhang2018_RDN}}
	\label{fig:RDN architecture}
	\vspace{\vspaceImage}
	\vspace{1mm}
\end{figure}

The overall RDN structure is depicted in Fig.~\ref{fig:RDN architecture}. In the beginning, low-level features are extracted from the LR input image~$\tilde{X}_\mathrm{LR,RGB}$ by the first two convolutional layers which results in a feature map with depth $G_0$. Afterwards, $D$ RDBs (red color) extract local hierarchical features by convolving and concatenating feature maps. 

\begin{figure}[t]
	\centering
	\includegraphics[width=1\linewidth]{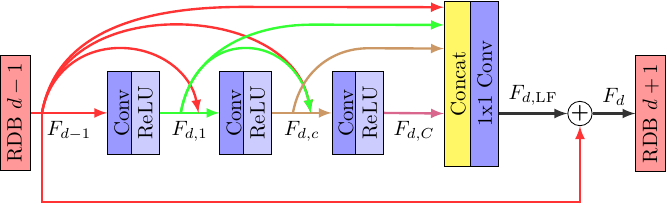}
	\caption{$d^\mathrm{th}$ RDB structure to extract local features; for simplicity, the concatenation layers between the convolutional layers are omitted.}
	\label{fig:RDB structure}
	\vspace{\vspaceImage}
\end{figure}

In Fig. \ref{fig:RDB structure} the structure of the $d^\mathrm{th}$ RDB is shown. There, $C$ convolutional layers with subsequent ReLU activation are used, which are fed with the output of the previous RDB $F_{d-1}$. Additionally, the outputs of the former convolutional layers are concatenated with $F_{d-1}$ and fed into the subsequent convolutional layers. The output of each convolutional layer has a depth of $G$, which is referred to as ``growth rate" in the original paper. After concatenating the outputs of all convolutional layers $F_{d, (1,c,...C)}$, the feature map depth is $G_0 + C\times G$. With the subsequent $1\times1$ convolutional layer, the output feature map depth $F_{d,LF}$ is transformed to $G_0$ again, in order to be able to add this low feature map to the first feature map $F_{d-1}$ and use residual learning. 

In the third step of RDN shown in Fig.~\ref{fig:RDN architecture}, called dense feature fusion, the outputs of all RDBs are first concatenated (yellow box) and then the feature map depth is reduced to a size of $G_0$ in order to add the resulting global features $F_\mathrm{GF}$ with the low-level feature map $F_{-1}$. This process derives global features by fusing the local features extracted previously by the RDBs. As the last step, the spatial resolution of the fused feature map is enhanced by the Efficient Sup-pixel CNN (ESPCNN) as presented in~\cite{shi2016}. Eventually, the resulting HR feature map is fed into a convolutional layer which results in an HR image~$\tilde{X}_\mathrm{HR,RGB}$ with the three RGB channels.


Details on the used implementation and how RDN is fit into the coding chain, are described in Sec. \ref{subsec: upscaling with RDN}.

\section{Coding Chain and Modifications}
\label{sec:coding chain}

The proposed coding chain is shown in Fig. \ref{fig:coding chain}. In this paper, the coding chain is proposed for VVC as encoder/decoder pair and VDSR or RDN as upscaling algorithm. In the following subsections the used implementations for the coding chain and its modifications are described.

\subsection{Downscaling}

First, the input sequence $X_\mathrm{orig}$ is spatially downscaled. In the scope of this paper, we always consider 4K sequences in YCbCr format with 4:2:0 color sub-sampling and a scaling factor $s$ of two, which results in a downscaled Y-channel $X_\mathrm{LR,Y}$ with full HD resolution ($1920\times1080$ pixels). Therefore, the chroma components have a resolution of $960\times540$ pixels. For the downscaling, every channel of each frame of the input sequence is downscaled by a factor of two with bicubic downscaling like it is proposed for VDSR and RDN.

\subsection{Encoding/Decoding}

After downscaling, the frame is encoded with the VVC reference software VVC Test Model (VTM) \cite{chen2019}. This encoder is set to compress the frames with fixed quantization parameter~(QP) values. For the conventional coding chain, the eleven QPs are set between 28 and 48 ($QP_\mathrm{conv}$) in steps of two. This differs from the JVET QP recommendations \cite{bossen2019}, since in the scope of this paper, low-bitrate scenarios are evaluated. In order to create rate-distortion curves for the coding chain with spatial down- and upscaling that are located in the same bitrate range, eleven QPs from 20 to 40  ($QP_\mathrm{scaled}$) are used for the coding chain with spatial up- and downscaling. As a result of the encoding process, the size of the resulting bitstream $b_\mathrm{LR}$ is measured and saved. Subsequently, the corresponding VTM decoder reconstructs the frame $\tilde{X}_\mathrm{LR}$, which is spatially downsampled and degraded by compression. 

\subsection{Upscaling with VDSR}
\label{subsec: upscaling with VDSR}

Since the decoded Y-channel of the frame has only full HD resolution, VDSR is investigated to increase the resolution back to the original $3840\times2160$ pixels. For that purpose, the frame is first bicubically upscaled. Afterwards, the self-implemented VDSR Tensorflow model is applied in order to regain the high frequency components. This Tensorflow model is taken instead of the reference VDSR implementation because of the possibility to self-train the model for high-resolution content and to increase the inference speed.

In comparison to the original VDSR, the used Tensorflow VDSR is trained on the \textit{DIV2K} dataset \cite{timofte2018} instead of the \textit{291} dataset that is originally used and which contains 291 natural low resolution images with a size below $500\times500$ pixels. This \textit{DIV2K} dataset is suggested by JVET for training neural nets in combination with the VVC \cite{li2019} and contains 800 training images with high resolution above 1.5 megapixels.

Since the \textit{DIV2K} training images have a higher resolution than the ones from set \textit{291}, data augmentation is waived for \textit{DIV2K} to get similar numbers of patches for both datasets. Additionally, only every sixteenth patch is considered in order to be able to take patches from every image of the \textit{DIV2K} dataset, and still keep the computational and hardware costs reasonable. Besides, the Adam optimizer~\cite{kingma2015} with a learning rate of 0.0001 is taken for training instead of stochastic gradient descent. Every 20 epochs, this learning rate is decreased by a factor of ten. Furthermore, the network is exclusively trained for a scaling factor of two. The other hyper-parameters are set as described for the original VDSR~\cite{kim2016}.

\def\widthPlots{0.95}
\begin{figure}[t]
	\centering
	\includegraphics[width=\widthPlots\linewidth, page=2]{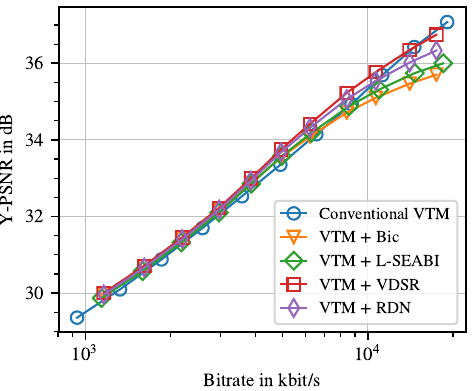}
	\caption{PSNR over bitrate for \textit{Campfire} sequence.}
	\label{fig:results psnr}
	\vspace{\vspaceImage}
\end{figure}

\subsection{Upscaling with RDN}
\label{subsec: upscaling with RDN}
For running RDN to upscale the coded frame with reduced spatial resolution, we use the Keras implementation and the pre-trained weights as published in~\cite{cardinale2018isr}. The used RDN model is implemented with $D=20$ RDBs, which each contain $C=6$ convolutional layers. Additionally, the feature map depth resulting from each RDB is set to $G_0=64$, as well as the depth $G$ of each feature map resulting from a convolutional layer inside a RDB. Similar to our VDSR implementation, the used RDN model is already trained on the \textit{DIV2K} dataset. Besides, the network is trained on maximizing the PSNR between LR and HR patch.

As already mentioned in Sec. \ref{subsec:RDN}, RDN takes an RGB image and returns an RGB image as output. In a first approach we converted each coded YCbCr frame in 4:2:0 subsampling format into an RGB image to feed the RDN model. Subsequently, a back-conversion to YCbCr has to be made, in order to compare the resulting upscaled frame with the original frame. But due to the color sub-sampling in the chroma channels, the edges are not preserved as sharp when converting the input to RGB. Hence, RDN cannot restore that many details and neither a decent PSNR nor VMAF can be achieved. Thus, we feed each $\tilde{X}_\mathrm{LR, RGB}$ input channel with the coded Y-channel $\tilde{X}_\mathrm{LR, Y}$ as second approach in order to maintain the high frequency components in the input. The resulting three output channels $\tilde{X}_\mathrm{HR, RGB}$ are converted to the Y output channel $\tilde{X}_\mathrm{HR, Y}$ according to Rec.ITU-R BT.601-7 \cite{itu_color2011} with

\begin{equation}
\tilde{X}_\mathrm{HR, Y} = 0.299 * Y_\mathrm{HR, R} + 0.587 * Y_\mathrm{HR, G} + 0.114 * Y_\mathrm{HR, B}.
\end{equation}

Besides, the two color channels $\tilde{X}_\mathrm{HR, Cb}$ and $\tilde{X}_\mathrm{HR, Cr}$ are upscaled with bicubic interpolation as it is done for VDSR. Although RDN not being trained for this procedure, it shows superior results than the first variant and is thus used for the subsequent experiments.

\section{Experimental Results}
\label{sec:experimental results}

\subsection{Simulation Setup}

For our simulation framework we use the VTM version~5.0~\cite{chen2019}. As test sequences, the first 30 frames of the eleven JVET sequences~\cite{bossen2019}, which contain natural 4K content, are simulated. Besides, the random-access test settings for 8 bit sequences from~\cite{bossen2019} are taken and show that the proposed framework works for I-, P-, and B-Frames.

As upscaling algorithms our own VDSR Tensorflow implementation is taken, the modified RDN implementation as described in Section \ref{subsec: upscaling with RDN}, as well as the bicubic interpolation and L\nobreakdash-SEABI reference implementation. Since bicubic, VDSR, and RDN results are derived from the same bitstream they have the same bitrate, whereas the L-SEABI bitstream is slightly different, since L-SEABI requires a $5\times5$ Gaussian filter before downscaling as proposed in \cite{georgis2016}, which results in a different $X_\mathrm{LR}$ and thus another $b_\mathrm{LR}$.

In the final evaluation, the resulting coded frames are objectively evaluated with the well-known PSNR metric in the Y-channel and the full-reference metric VMAF\cite{li2016_vmaf}, which combines multiple existing quality metrics and parses them into a support vector machine to model human subjective perception. VMAF returns a quality between 0 and 100, with 0 being the worst, and 100 being the best quality. The VMAF implementation is taken from \cite{vmaf_git} and is used with the default VMAF model \textit{vmaf\_v0.6.1.pkl}. We also measured the coded sequences with the VMAF 4K model, but the relative differences between the measured curves only changed marginally such that we focus on the default model.

\begin{figure}[t]
	\centering
	\includegraphics[width=\widthPlots\linewidth, page=2]{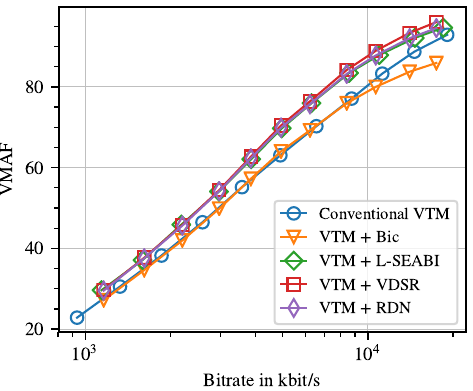}
	\caption{VMAF over bitrate for \textit{Campfire} sequence.}
	\label{fig:results vmaf}
	\vspace{\vspaceImage}
\end{figure}
\begin{table*}[t]
	\caption{BDR Savings in \% for two QP ranges using conventional VTM coding chain as anchor; Please note that a positive value corresponds to a BD-rate saving compared to the anchor. The best values are set in bold numbers. Additionally, the frames per second (FPS) for each sequence are given.}
	\label{tab:bd rate}
	\centering
	\resizebox{\textwidth}{!}{%
		\begin{tabular}{lc||rrrr|rrrr||rrrr|rrrr}
			& \multicolumn{1}{l| |}{}    & \multicolumn{8}{c| |}{\begin{tabular}[c]{@{}c@{}}$QP_\mathrm{scaled}=\{34, 36, 38, 40\}$,\\ $QP_\mathrm{conv}=\{42, 44, 46, 48\}$\end{tabular}}              & \multicolumn{8}{c}{\begin{tabular}[c]{@{}c@{}}$QP_\mathrm{scaled}=\{26, 28, 30, 32\}$,\\ $QP_\mathrm{conv}=\{34, 36, 38, 40\}$\end{tabular}}                \\
			& \multicolumn{1}{l| |}{}& \multicolumn{4}{c|}{Y-PSNR BD-Savings}        & \multicolumn{4}{c| |}{VMAF BD-Savings}    & \multicolumn{4}{c|}{Y-PSNR BD-Savings} & \multicolumn{4}{c}{VMAF BD-Savings} \\
			& \multicolumn{1}{l| |}{}    &               &               &               &         &               &               &              &              &          &             &             &          \\
			& \multicolumn{1}{l| |}{}    &               &               &               &         &               &               &              &              &          &             &             &          \\
			& \multicolumn{1}{l| |}{}    &               &               &               &         &               &               &              &              &          &             &             &          \\
			\multicolumn{1}{l}{Sequence Name}    & \multicolumn{1}{c| |}{FPS} & {\tabrotate{Bicubic}} &  \tabrotate{L-SEABI} &     \tabrotate{VDSR} & \tabrotate{RDN} &  \tabrotate{Bicubic} &  \tabrotate{L-SEABI} &     \tabrotate{VDSR}& \tabrotate{RDN} & {\tabrotate{Bicubic}} &  \tabrotate{L-SEABI} &     \tabrotate{VDSR}& \tabrotate{RDN} &  \tabrotate{Bicubic} &  \tabrotate{L-SEABI} &     \tabrotate{VDSR}  &   \tabrotate{RDN} \\ \hline
			\textit{BuildingHall2}    & 50                       & 6.3           & 13.5          & 14.0          & \textbf{14.1} & -4.5    & \textbf{22.1} & 19.8          & 22.0          & -22.3        & -7.2          & -6.1  & \textbf{-5.2}  & -30.3   & 11.0         & 9.8           & \textbf{13.5} \\
			\textit{Campfire}         & 30                       & 6.3           & 2.7           & \textbf{7.3}  & 5.6           & -1.5    & 12.6          & \textbf{13.1} & 11.9          & 3.9          & 5.4           & \textbf{14.3}  & 11.4  & 1.6     & 28.7         & \textbf{33.0} & 30.0          \\
			\textit{CatRobot1}        & 60                       & 4.7           & 5.9           & \textbf{6.5}  & 4.8           & -3.7    & 7.6           & 7.6           & \textbf{9.4}  & -15.8        & -11.8         & \textbf{-5.2}  & -11.7 & -20.8   & 0.1          & 1.7           & \textbf{2.7}  \\
			\textit{DaylightRoad2}    & 60                       & 5.3           & 6.1           & \textbf{6.8}  & 6.2           & -4.3    & 6.3           & 6.5           & \textbf{7.8}  & -21.4        & -19.2         & \textbf{-13.8} & -16.6 & -23.8   & -5.0         & -1.1          & \textbf{-0.1} \\
			\textit{Drums}            & 100                      & 5.6           & \textbf{6.3}  & 6.2           & 4.5           & -2.6    & \textbf{9.2}  & 6.8           & 8.6           & -11.0        & \textbf{-6.1} & -6.5           & -10.9 & -17.8   & \textbf{3.9} & 1.2           & 3.7           \\
			\textit{FoodMarket4}      & 60                       & 16.7          & \textbf{17.4} & 15.8          & 13.6          & 35.5    & 32.6          & 36.9          & \textbf{39.5} & 11.7         & \textbf{12.0} & 9.8            & 1.5   & 16.8    & 16.3         & 19.5          & \textbf{23.0} \\
			\textit{ParkRunning3}     & 50                       & 20.9          & 19.1          & \textbf{21.7} & 20.5          & 16.0    & 27.9          & 27.9          & \textbf{28.5} & 11.9         & 12.0          & \textbf{18.5}  & 15.8  & 10.9    & 29.0         & 31.6          & \textbf{32.0} \\
			\textit{RollerCoaster2}   & 60                       & \textbf{18.9} & 18.4          & 18.4          & 17.4          & 26.5    & 27.1          & 28.4          & \textbf{29.6} & \textbf{8.5} & 7.2           & 7.7            & 4.4   & 8.7     & 12.1         & 13.0          & \textbf{15.0} \\
			\textit{Tango2}           & 60                       & \textbf{9.4}  & 9.1           & 8.9           & 6.6           & 17.0    & 16.8          & 19.4          & \textbf{20.7} & \textbf{4.6} & 3.7           & 4.1            & -2.1  & 2.4     & 6.7          & 7.4           & \textbf{9.3}  \\
			\textit{ToddlerFountain2} & 60                       & \textbf{-0.3} & \textbf{-0.3} & -0.4          & -1.0          & 3.3     & 7.8           & 7.9           & \textbf{9.0}  & -6.4         & -6.4          & \textbf{-5.1}  & -6.4  & -3.8    & 3.8          & 4.0           & \textbf{4.5}  \\
			\textit{TrafficFlow}      & 30                       & 8.0           & 9.4           & \textbf{10.4} & 7.4           & -2.0    & \textbf{12.0} & 9.3           & 10.5          & -17.8        & -13.4         & \textbf{-7.8}  & -16.2 & -24.4   & \textbf{3.0} & -0.1          & 1.4           \\ \hline
			Average overall  & \multicolumn{1}{l||}{}    & 9.3           & 9.8           & \textbf{10.5} & 9.1           & 7.2     & 16.6          & 16.7          & \textbf{18.0} & -4.9         & -2.2          & \textbf{0.9}   & -3.3  & -7.3    & 10.0         & 11.0          & \textbf{12.3}
	\end{tabular}}
\vspace{-4mm}
\end{table*}

\subsection{Evaluation}

In order to evaluate the proposed framework, the five RD-curves for the conventional VTM coding chain (blue), VTM with bicubic interpolation (orange), VTM with L-SEABI upscaling (green), VTM with VDSR  upscaling (red), and VTM with RDN (purple) are plotted exemplary in Fig. \ref{fig:results psnr} for the \textit{Campfire} sequence with PSNR as quality metric. All four measured coding chains with spatial up- and downscaling show superior results than the conventional VTM up to a certain bitrate. For the coding chain with simple bicubic upscaling, this critical bitrate is around 7000 kbit/s. Using L-SEABI upscaling, the critical bitrate is around 9000 kbit/s. By increasing the spatial resolution with VDSR and RDN, the critical bitrate can be raised to 14500 kbit/s and 11000 kbit/s, respectively. When considering VMAF as in Fig. \ref{fig:results vmaf}, the critical bitrate is even higher. The RD-curves for L-SEABI, VDSR, and RDN do not cross the curve for conventional coding before 19000 kbit/s. For the other sequences a similar behavior can be seen, but the critical bitrates vary from 1000 to above 40000 kbit/s depending on the framerate and the sequence content.

For a better comparison at two different quality areas, we list the Bj\o ntegaard delta rate~(BDR)~\cite{bjontegaard2001} in terms of bitrate savings in Table~\ref{tab:bd rate} with the conventional VTM coding chain as anchor. BDR calculates the bitrate savings for two given RD-curves at the same quality. Originally, BDR is proposed for PSNR, however, we also show BDR values where we replaced the PSNR metric with VMAF.

Considering a very low video quality ($QP_\mathrm{conv}=\{42, 44, 46, 48\}$), coding with the investigated coding chain with spatial downscaling results in BDR savings with respect to PSNR above 9~\%. Upscaling the coded frames with VDSR results in the highest bitrate savings of 10.5~\%. Evaluating BDR with respect to VMAF shows even higher coding gains up to 18.0~\% for RDN as upscaling algorithm. The discrepancy for RDN between PSNR and VMAF is possibly caused by feeding the network, which was originally trained for RGB data, with only the Y-channel. Thus, the exact values cannot be reproduced, but high frequencies and edges can be restored well, which has a negative impact on PSNR but not on VMAF. Additionally, the bicubic interpolation is penalized for delivering blurry frames considering VMAF. At best, the investigated coding chain with RDN can save 39.5~\% for the \textit{FoodMarket4} sequence.

The investigated coding chain also works when using lower QP values, albeit the bitrate savings compared to conventional VVC are lower on average and even result in negative BDR values with respect to PSNR except when using VDSR. Considering VMAF for BDR still results in massive bitrate savings of up to 12.3~\%. In average, RDN outperforms VDSR and L-SEABI by 1.0~\% and 2.3~\%, respectively.

\subsection{Visual Results}
Additionally, visual results are shown for a zoomed area of the second frame of the \textit{Campfire} sequence and a bitrate around 6300 kbit/s in Fig. \ref{fig:visual results campfire}. It can be observed that the coding chain with spatial down- and upscaling retains more details that are lost by coding at the original resolution. For the area that is compressed with the conventional coding chain, a spark is missing at the bottom left corner, which is still present in the frames that are coded with the proposed coding chain, except for L-SEABI upscaling. Besides, blocking artifacts can be noticed for conventional coding on the rightmost spark in the image. Furthermore, the sparks in the frame that is bicubically interpolated appear more blurry than for interpolating the frame with VDSR or RDN. The results of the two machine-learning-based super-resolution algorithms are visually comparable.

\subsection{Complexity Evaluation}

For a possible real-time application the runtime of the resolution enhancing algorithms is decisive. However, it is hard to compare these algorithms properly since VDSR and RDN are executed on a GPU, whereas L-SEABI runs on the CPU and the used implementation is explicitly not intended for comparing runtime. Nevertheless, we can say that running VDSR takes around 1 second to upscale the Y channel of a full HD frame to 4K resolution on a NVIDIA GeForce RTX 2080 Ti. On the same unit, RDN takes between 6 and 8 seconds. The L-SEABI takes around 1 second on an Intel Xeon E3-1275 v6 @ 3.8 GHz in the proposed coding chain for upscaling the Y-channel. The Cb and Cr channels take additional 0.5 seconds. This makes all upscaling algorithms not feasible for possible real-time applications on current hardware. However, these runtimes are significantly lower than the encoding time of VTM 5.0 which lies above one minute per 4K frame and a QP of 48.

\def\imageWidth{0.23}
\def\yOffset{-5mm}
\begin{figure}[t] 
	\centering
	\begin{subfigure}{\imageWidth\textwidth}
		\includegraphics[width=\linewidth]{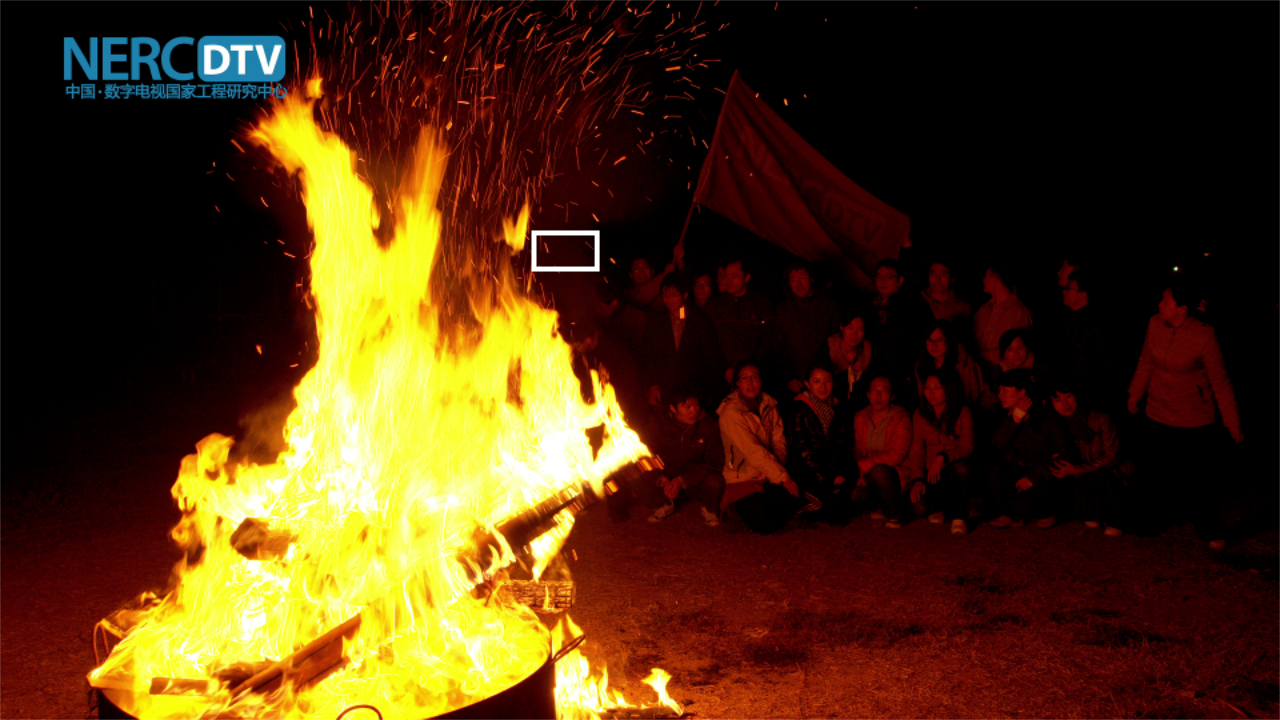}
		\vspace{\yOffset}
		\caption{Complete sequence} \label{fig:a}
	\end{subfigure}
	
	\begin{subfigure}{\imageWidth\textwidth}

		\includegraphics[width=\linewidth]{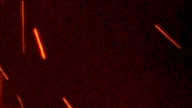}
		\vspace{\yOffset}
		\caption{Uncompressed patch of size $192\times108$ pixels} \label{fig:b}
	\end{subfigure}
	\begin{subfigure}{\imageWidth\textwidth}

		\includegraphics[width=\linewidth]{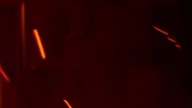}
		\vspace{\yOffset}
		\caption{Conventional VTM \\ @ 6570 kbit/s} \label{fig:c}
	\end{subfigure}
	\hspace*{\fill}
	
	\begin{subfigure}{\imageWidth\textwidth}
		\includegraphics[width=\linewidth]{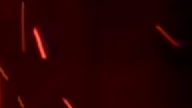}
		\vspace{\yOffset}
		\caption{VTM with bicubic upscaling\\ @ 6236 kbit/s} \label{fig:d}
	\end{subfigure}
	\begin{subfigure}{\imageWidth\textwidth}
		\includegraphics[width=\linewidth]{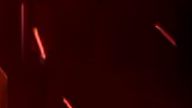}
		\vspace{\yOffset}
		\caption{VTM with L-SEABI upscaling\\ @ 6314 kbit/s} \label{fig:e}
	\end{subfigure}
	\hspace*{\fill}
	
	\begin{subfigure}{\imageWidth\textwidth}
		\includegraphics[width=\linewidth]{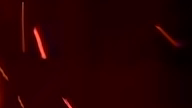}
		\vspace{\yOffset}
		\caption{VTM with VDSR upscaling \\ @ 6236 kbit/s} \label{fig:f}
	\end{subfigure}
	\begin{subfigure}{\imageWidth\textwidth}
		\includegraphics[width=\linewidth]{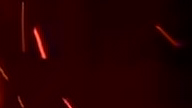}
		\vspace{\yOffset}
		\caption{VTM with RDN upscaling \\ @ 6236 kbit/s} \label{fig:g}
	\end{subfigure}
	\hspace*{\fill}
	
	\caption{Visual results for second frame of \textit{Campfire} sequence; $QP_\mathrm{scaled}=28$, $QP_\mathrm{conv}=36$.}
	\label{fig:visual results campfire}
	\vspace{\vspaceImage}
\end{figure}

\section{Conclusion}
\label{sec:conclusion}

In this paper, a coding chain was presented that combines next-generation video coding with two machine learning based networks for upscaling named VDSR and RDN for 4K data. Various results showed that the coding chains with the machine learning based super-resolution algorithms achieve a superior rate-distortion behavior for 4K data compared to conventional coding at high QP values. Besides, the visual results look more appealing and artifacts can be reduced. However, one still has to take care of the runtime in a possible real-time coding scenario. There, it might not be feasible to run VDSR and RDN without significantly reducing their layers. Additionally, it can be considered to upscale the coded frame with bicubic interpolation, which also achieved significant coding gains compared to conventional VVC, and then only enhance the quality with VDSR in regions with high frequency content where it is beneficial. Moreover, the machine learning based networks can be trained on images with compression artifacts to further increase the rate-distortion performance at low bitrates.



\vspace{-2mm}

\bibliographystyle{IEEEtran}
\bibliography{/home/fischer/Paper/jabref_literature_research_ms2.bib}

\end{document}